\documentclass{aa}  
\usepackage{graphicx}
\usepackage{txfonts}
\begin{document}
   \title{No unique quadratic ephemeris for VW Cep}
   
   \subtitle{RLOF-reversals in a shallow contact binary}

   \author{W. Van Rensbergen, J.P. De Greve \and N. Mennekens}
 
  \institute{Astrophysical Institute, Vrije Universiteit Brussel, Pleinlaan 2, 1050 Brussels, Belgium\\
   \email {wvanrens@vub.ac.be}
   }
              
   \date{Received June 6, 2013; accepted}

% \abstract{}{}{}{}{} 
% 5 {} token are mandatory
 
  \abstract
  % context heading (optional)
  % {} leave it empty if necessary  
   {From the eclipsing binary VW Cep, 1963 epochs of primary and secondary minima have been registered since 1926.}
  % aims heading (mandatory)
   {Because VW Cep is a shallow contact binary the direction of the flow from one component to the other could possibly be reversed. We aim to fit the observed eclipses into a scenario in which reversals may occur.}
  % methods heading (mandatory)
   {Orbital period changes can not be fitted into a unique quadratic ephemeris, because of discontinuities in the (O-C) vs E graph. We try to obtain better agreement through the allowance of RLOF reversal in this binary system.}
  % results heading (mandatory)
   {From 1926 until 1943 the orbital period increase fits into a model of mass transfer from the less massive component. From 1946 on the direction of the mass transfer changed so that the most massive component became the donor star, as shown by a systematic period decrease. A sudden period increase in 1999 introduces a currently underway episode of mass transfer from the less massive component.}
  % conclusions heading (optional), leave it empty if necessary 
   {The ephemeris of VW Cep can be fitted into a model in which the direction of the mass transfer was reversed twice since 1926.}
   
   
    \authorrunning{W. Van Rensbergen et al.}  
     \titlerunning{Ephemeris of VW Cep}
   \maketitle
%________________________________________________________________

\section{Introduction}

\begin{figure*}[!ht]
\centering
\includegraphics[width=9.6cm]{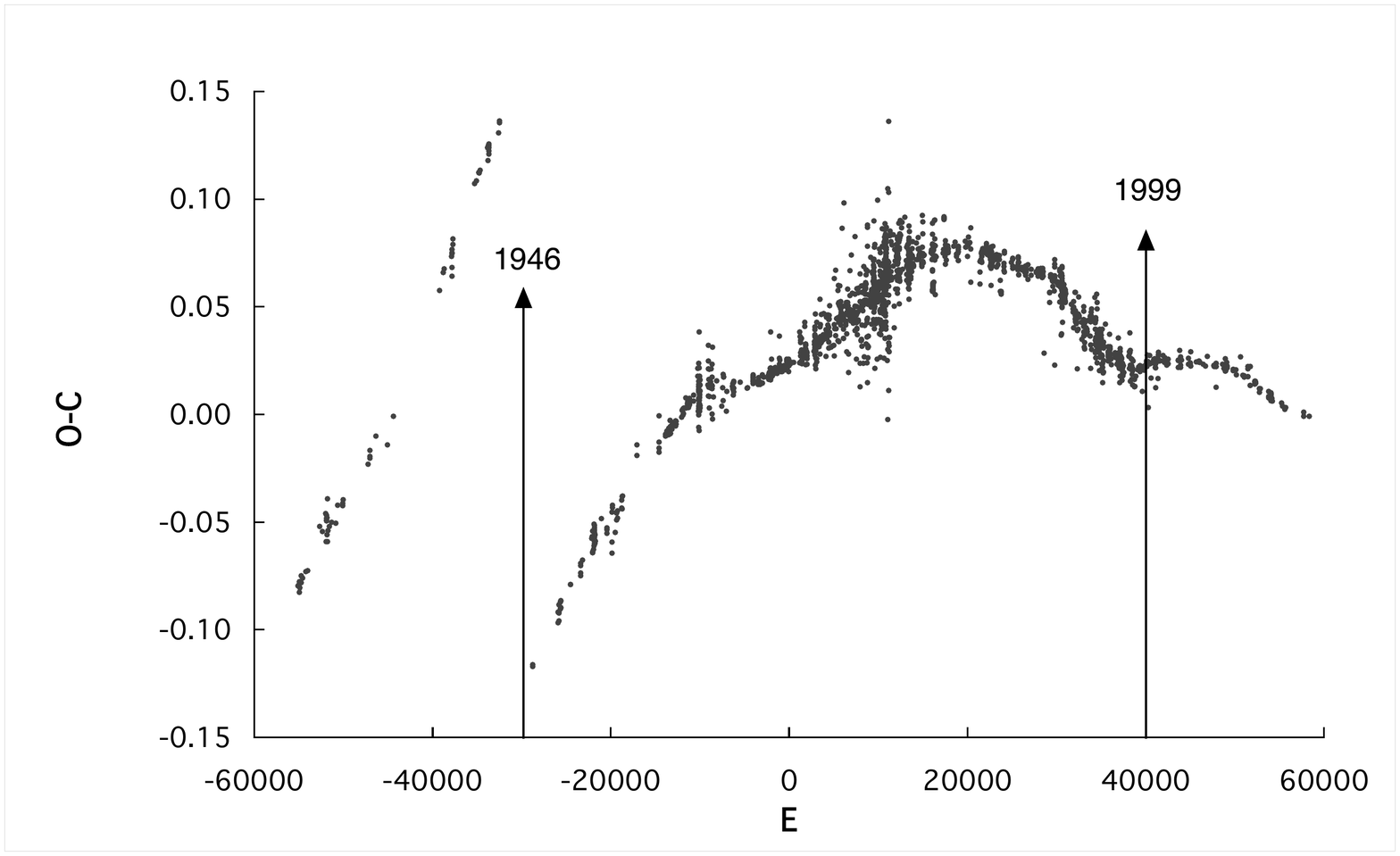}
\caption{VW Cep shows mass transfer from its most massive to its less massive component between 1946 and 1999. Before 1946 and after 1999 clearly something happened.} 
\label{fig_fig1}
\end{figure*}

A plot of (O-C) as a function of E is a powerful tool to determine mass transfer in a binary. The absence of mass transfer implies that the epochs of light minima caused by the eclipses follow a linear ephemeris. When mass transfer is at work a quadratic ephemeris is needed. A positive value of the coefficient A2 of $E^{2}$ reveals a period increase caused by mass transfer from the less massive donor to the more massive gainer. When the donor is the more massive component,  one finds a negative value of the coefficient of A2 implying a period decrease.

Fig. {\ref{fig_fig1}} shows 1963 values of (O-C) as a function of E. The observed values (O) (between 1926 and 2012) are obtained from the (O-C) gateway made available on internet by Paschke \& Br$\grave{a}$t (\cite{Paschke}) \footnote{http://astro.sci.muni.cz/variables/ocgate/}.
The computed values (C) are obtained from a linear ephemeris with a constant orbital period of $\approx$ 0.2783 days. Between 1946 and 1999 the (O-C) values follow a well defined downwards turned parabola. The quadratic term is negative and mass is transferred from the more to the less massive component. The necessary continuity of the curve is however clearly broken before 1946 and after 1999.

\section{Physical parameters}

Table {\ref{tab_tab1}} lists values of mass, radius and effective temperatures of the components obtained and used by various authors. Pustylnik \& Niarchos (\cite{Pustylnik}) agree upon a most massive component of $\approx$ 0.9 $M_{\odot}$ with a radius of $\approx$ 0.93 $R_{\odot}$, orbiting with a less massive companion having $\approx$ 0.25 $M_{\odot}$ with a radius of $\approx$ 0.5 $R_{\odot}$. The most massive component is the coolest with $T_{1,eff}$ $\approx$ 5000~K having a slightly hotter companion with  $T_{2,eff}$ $\approx$ 5200~K.

Kasz$\acute{a}$s et al. (\cite{Kaszas}) find a somewhat more massive binary with $M_{1}$ $\approx$ 1.015 $M_{\odot}$ and $M_{2}$ $\approx$ 0.355 $M_{\odot}$. The data of Hendry \& Mochnacki (\cite{Hendry}) have been omitted from Table {\ref{tab_tab1}}  because their small radii produce VW Cep as a detached binary in which mass-transfer is not possible.

The HRD positions that can be drawn from Table {\ref{tab_tab1}}  show a binary containing a slightly underluminous star with a largely overluminous less massive companion. Both on the Main Sequence. How such a configuration can be obtained needs further modelling of binary evolution.

We assume that the coefficient A2 of the quadratic ephemeris determines the mass transfer rate. 

Other influences are considered to be of minor influence on the orbital evolution of the binary. For instance the well established presence of a tertiary component, already mentioned by Popper (\cite{Popper} with a mass  of $\approx$ 0.58 $M_{\odot}$, recalculated as $\approx$ 0.74 $M_{\odot}$ with an orbital period of  $\approx$ 30 y by Zasche \& Wolf (\cite{Zasche}.

The mass transfer rate is however not very constant.  Binnendijk (\cite{Binnendijk}) mentions a period change that occurred around 1945. Hill (\cite{Hill}) calculates three major period changes in 1944, 1950 and 1961, repectively. Pribulla et al. (\cite{Pribulla}) mention a sudden period increase in 1999. 

\vspace{1mm}

When $\dot{P}$ (in $\frac{d}{y}$) was measured over a sufficient number of orbital periods, the mass transfer rate $\dot{M}_{d}$ (in $\frac{M_{\odot}}{y}$) calculated with the conservative assumption (given by equation {\ref{Mdot}})\footnote{subscript $g$ for gainer and $d$ for donor} is also included in Table {\ref{tab_tab1}}. Being the mass loss rate of the donor star, the quantity $\dot{M}_{d}$  is negative.

\begin{equation}
\dot{M}_{d}~ \left (\frac{1}{M_{g}} -  \frac{1}{M_{d}}\right )~=~\frac{1}{3~P}~\dot{P},
\label{Mdot}
\end{equation}

With an orbital period $P$ $\approx$ 0.2783 days both stars in the binary have radii very close or a little larger than their Roche radii. VW Cep is thus a shallow contact binary. 

\begin{table*}
\caption{Physical quantities related to VW Cep.
Columns: (1) Popper (\cite{Popper}), (2) Binnendijk (\cite{Binnendijk}), (3) Hill (\cite{Hill}), (4)  Pribulla et al. (\cite{Pribulla}), (5) Kasz$\acute{a}$s et al. (\cite{Kaszas})}
\label{table:Ktabel}
\centering
\begin{tabular}{c c c c c c c c}
\hline
& & & & & \\
 & $1$ & $2$ & $3$ & $4$ & $5$\\
\hline
$M_1$ $sin^3$~i & 0.62 & 0.61 & 0.668 & 0.767 & 0.767\\
$M_2$ $sin^3$~i  & 0.2 & 0.25 & 0.182 & 0.268 & 0.268\\
i&$56^o$&$63^o$&$65^o$&$65^o$.6&$65^o$.6\\
$M_1$&1.1&0.87&0.897&1.015&1.015\\
$M_2$&0.35&0.35&0.247&0.355&0.355\\
$R_1$ & 0.6 & 0.939 & 0.93 & 0.94 & 0.94\\
$R_2$& 0.38 & 0.53 & 0.5 & 0.58 & 0.58\\
$T_{1,eff}$& 4700 &  & 4960 & 5050 & 5050 \\
$T_{2,eff}$& 5250 & & 5200 & 5444 & 5444\\
$Log~{L_{1}}$ & -0.8 & & -0.326 & -0.285 & -0.285\\
$Log~{L_{2}}$ &-1.0 & & -0.783 & -0.574 & -0.574\\
$\dot{P}$ &  & & -[1.586-4.425]~E-6&-2.163E-7 & -2.135E-7\\
$\dot{M}_{d}$ & & & -[6.476-18.06]~E-7 & -1,414E-7 & -1.396~E-7\\
\hline
\end{tabular}
\label{tab_tab1}
\end{table*}

\section{Two RLOF-reversals}

The masses and radii given by Hill (\cite{Hill}) and Pustylnik \& Niarchos (\cite{Pustylnik}) fit VW Cep very well into a shallow contact binary. Mass transfer rates are calculated with this model.

\subsection{The event between 1943 and 1946}

All the quantities in Table {\ref{tab_tab1}} are determined from data after 1946 and before 1999. In this era a plot of 1761 (O-C)-values vs E is convincingly described by a quadratic ephemeris with:
\vspace{1mm}

A2=-8.153E-11 or $\dot{P}$ = -2.140E-7 $\frac{d}{y}$, 

leading with relation ({\ref{Mdot}}) to a mass transfer rate $\dot{M}_{d}$ = -8.872E-8 $\frac{M_{\odot}}{y}$ from the most massive component to its less massive companion. With the stellar masses in Table {\ref{tab_tab1}} one obtains with equation {\ref{Mdot}} absolute values of mass transfer rates that are 50~$\%$ (Popper \cite{Popper}), 70~$\%$ (Binnendijk \cite{Binnendijk}) and  60~$\%$ (Kasz$\acute{a}$s et al. \cite{Kaszas}) larger.
\vspace{1mm}

As can be seen from Fig. {\ref{fig_fig1}} the plot of the first 64 (O-C)-values vs E (1926.4-1943.6) do not fit into that picture. An analysis of these 64 observed eclipses as a function of epoch leads however also to a clear parabola with quadratic term:
\vspace{1mm}
A2=+1.414E-10, leading with relation ({\ref{Mdot}}) to a mass transfer rate of -1.539E-7 $\frac{M_{\odot}}{y}$ from the less massive component to its more massive companion.
\vspace{1mm}

This indicates that the shallow contact binary underwent a RLOF-reversal between 1943.6 and 1946.5. From these three years no observations of epochs of light minima are available.

\begin{figure*}[!ht]
\centering
\includegraphics[width=9.6cm]{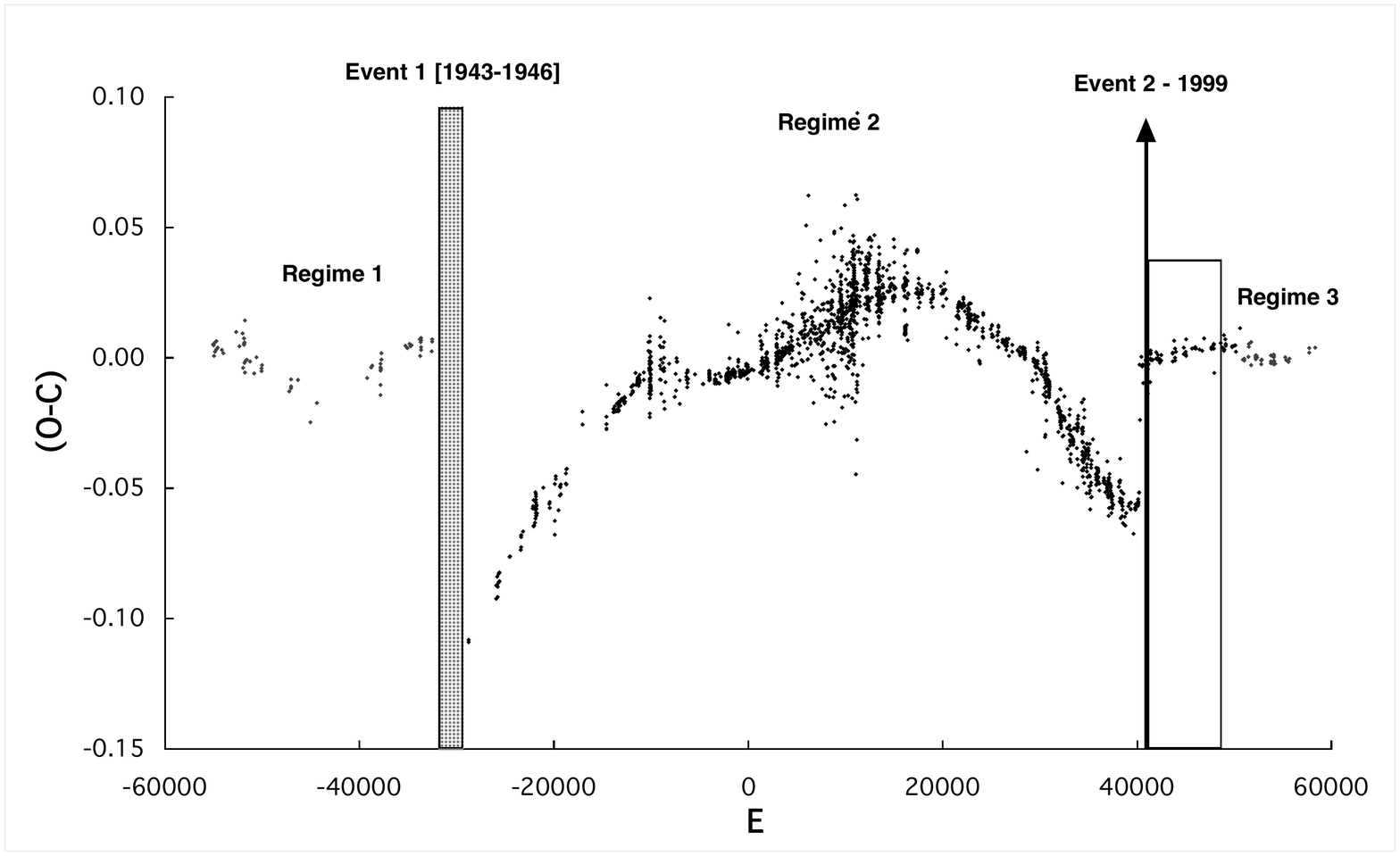}
\caption{Before 1943 the mass transfer was from the less to the more massive component. Between 1943 and 1946 the most massive took over the donor task. In 1999 the opposite effect occurred and the less massive star again became the donor star.} 
\label{fig_fig2}
\end{figure*}

The influence of this RLOF-reversal is shown in Fig. {\ref{fig_fig2}}. Comparison with Fig. {\ref{fig_fig1}} shows tat the oldest discontinuity has disappeared, due to a sudden small change of the orbital period as an consequence of a RLOF-reversal between 1943.5 and 1946.4.

\subsection{Consequences of the event of 1999}

Pribulla et al. (\cite{Pribulla}) mention that an episodic  mass transfer from the less to the more massive component might be at the origin of the sudden period increase detected in 1999. This announced second RLOF-reversal since 1926 is further investigated in this paper. The consequence of this event is visible at the right end of Fig. {\ref{fig_fig1}}. Only the 53 last observations of light minima as a function of epoch lead to a fairly clear parabola with quadratic term:
\vspace{1mm}
A2=+2.575E-10, leading with relation ({\ref{Mdot}}) to a mass transfer rate of -2.803E-7 $\frac{M_{\odot}}{y}$ from the less massive component to its more massive companion.
\vspace{1mm}

\begin{figure*}[!ht]
\centering
\includegraphics[width=9.6cm]{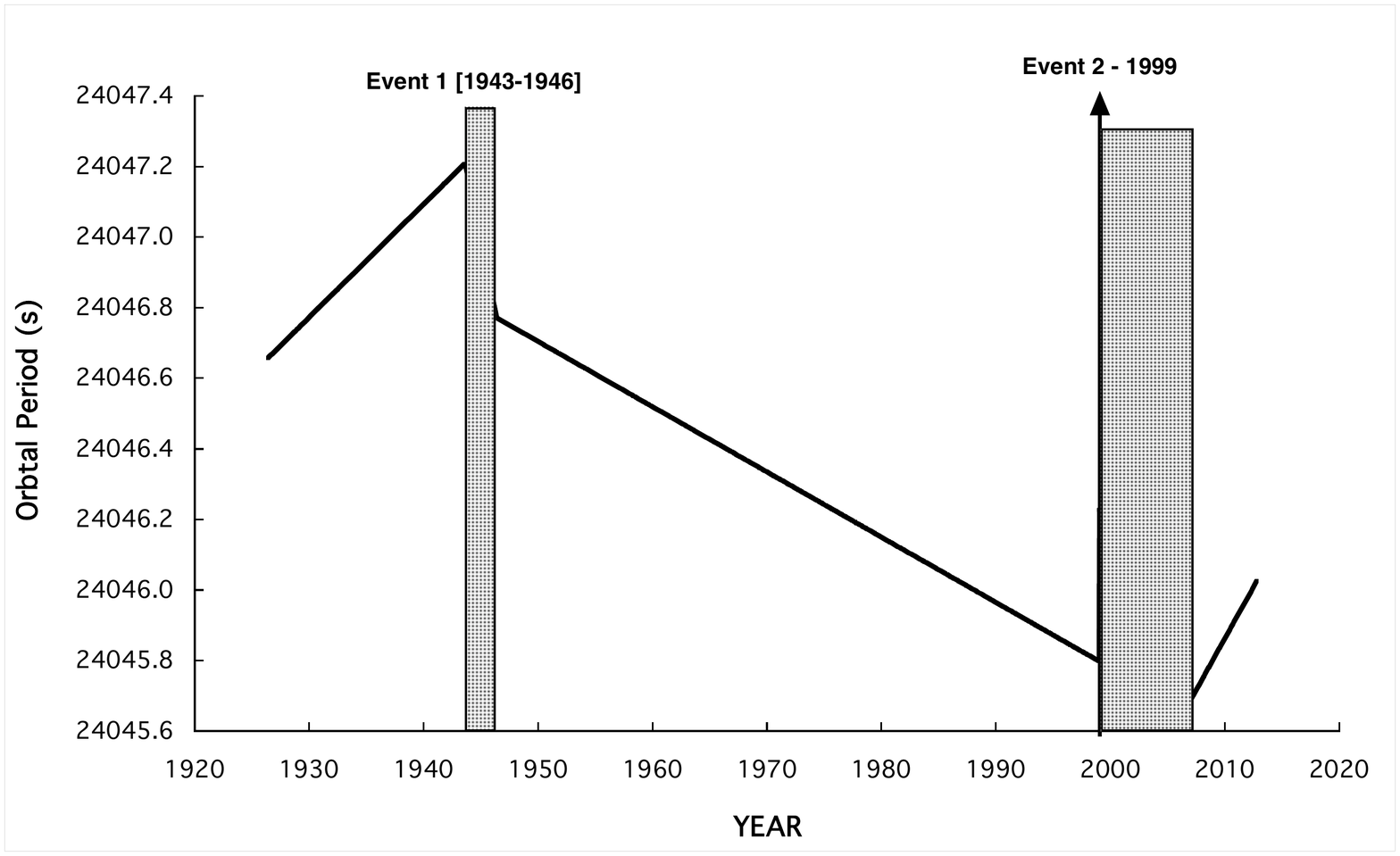}
\caption{The orbital period varies over  of $\approx$ 1.5 seconds between 1926 and 2013, due to RLOF reversals around 1945 and 1999.}
\label{fig_fig3}
\end{figure*}

We conclude that the sudden period increase detected in 1999 leads to a gradual inverting of the RLOF from the more massive donor to the less massive donor star. The first 85 measured epochs of light minima reversal show a rather chaotic behaviour. This initial transition phase after 1999 is shown by an transparent rectangle in Fig. {\ref{fig_fig2}}. From 2007 on, a rather steady flow from less massive donor to more massive gainer star has been observed over a time interval covering 53 observed light minima.

The effect of the two RLOF-reversals on (O-C) vs E-graph is shown in Fig. {\ref{fig_fig2}}. 

The effect of the two RLOF-reversals on the orbital period is shown in Fig. {\ref{fig_fig3}}. The orbital period varies over an amount of $\approx$ 1.5 sec, so that an orbital period of $\approx$ 0.2783 d is always found in the calibration of a linear ephemeris.

\section{Conclusions}

Since 1926, 1963 epochs of light minima have been observed.

\begin{flushleft}
$\bullet$~~The first 64 eclipses measured until 1943.6 show a mass transfer rate of -1.539E-7 $\frac{M_{\odot}}{y}$ from the less massive component to its more massive companion. \\ 
$\bullet$~~No further eclipses have been observed until the middle of 1946.\\
$\bullet$~~The next 1761 eclipses measured until 1999 show a mass transfer rate of -8.872E-8 $\frac{M_{\odot}}{y}$ from the most massive component to its less massive companion. A RLOF-reversal had thus taken place between 1943.5 and 1946,5\\  
$\bullet$~~A sudden increase of the orbital period has been observed in 1999. The next 85 observed eclipses until 2007 do not show a clear RLOF-reversal.\\  
$\bullet$~~The last 53 eclipses measured until the end of 2012 show however a mass transfer rate of -2.803E-7 $\frac{M_{\odot}}{y}$ from the less massive component to its more massive companion. It has thus taken some time between 1999 and 2007 to stabilize this second RLOF-reversal since 1926.\\  
$\bullet$~~The last established RLOF-reversal yields the less accurate determination of the mass transfer rate. Further observations are needed to establish more accurately the further orbital evolution of VW Cep, e.g. confirming the recently re-established flow from less the massive donor to the more massive gainer.
\\  
\end{flushleft}

\end{document}